\documentclass{article}

\usepackage{arxiv}

\usepackage[utf8]{inputenc} % allow utf-8 input
\usepackage[T1]{fontenc}    % use 8-bit T1 fonts
\usepackage{hyperref}       % hyperlinks
\usepackage{url}            % simple URL typesetting
\usepackage{booktabs}       % professional-quality tables
\usepackage{amsfonts}       % blackboard math symbols
\usepackage{nicefrac}       % compact symbols for 1/2, etc.
\usepackage{microtype}      % microtypography
\usepackage{lipsum}		% Can be removed after putting your text content
\usepackage{graphicx}
\usepackage[square, numbers, sort&compress]{natbib}
\usepackage{doi}
\usepackage{amsmath}
\usepackage{listings}
\usepackage{xcolor}
\usepackage[ruled,vlined]{algorithm2e}

\lstdefinestyle{py}{
    language=Python,
    basicstyle=\ttfamily\small,
    keywordstyle=\color{blue!70!black}\bfseries,
    commentstyle=\color{gray}\itshape,
    stringstyle=\color{purple!70!black},
    showstringspaces=false,
    tabsize=4,
    breaklines=true,
    frame=single,
    captionpos=b
}

\SetKw{Or}{or}
\SetKw{And}{and}

\title{A 58-Addition, Rank-23 Scheme for General 3×3 Matrix Multiplication}

\author{\href{https://orcid.org/0000-0001-8047-0114}{\includegraphics[scale=0.06]{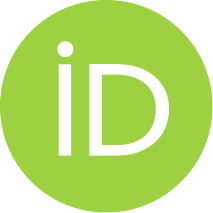}\hspace{1mm}Andrew I.~Perminov}\\
	Research Center for TAI\\
	Institute for System Programming\\
	Moscow \\
	\texttt{perminov@ispras.ru}
}

\renewcommand{\headeright}{}
\hypersetup{
pdftitle={A 58-Addition, Rank-23 Scheme for General 3×3 Matrix Multiplication},
pdfsubject={cs},
pdfauthor={Andrew I.~Perminov},
pdfkeywords={Fast matrix multiplication, Additive complexity, Ternary coefficients set},
}

\begin{document}
\maketitle

\begin{abstract}
This paper presents a new state-of-the-art algorithm for exact 3×3 matrix multiplication over general non-commutative rings, achieving a rank-23 scheme with only 58 scalar additions. This improves the previous best additive complexity of 60 additions~\citep{stapleton202560} without a change of basis. The result was discovered through an automated search combining ternary-restricted flip-graph exploration with greedy intersection reduction for common subexpression elimination. The resulting scheme uses only coefficients from $\{-1, 0, 1\}$, ensuring both efficiency and portability across arbitrary fields. The total scalar operation count is reduced from 83 to 81.
\end{abstract}

% keywords can be removed

\section{Introduction}
The computational complexity of matrix multiplication remains a fundamental problem in theoretical computer science and numerical linear algebra. Since Strassen's discovery of a sub-cubic time algorithm~\citep{strassen1969gaussian}, the search for practical, low-complexity schemes for small matrix formats has been a persistent research direction. For 3×3 matrices, the lowest known rank (in non-commutative case) is 23~\citep{Laderman1976ANA}. Any scheme of this rank requires 23 essential multiplications of linear combinations of input elements. However, the total number of scalar additions (and subtractions) needed to form these combinations is a crucial practical metric, as it directly impacts the efficiency of software implementations like the Basic Linear Algebra Subprograms (BLAS).

Significant effort has focused on reducing the additive complexity of rank-23 schemes. Laderman's original construction used 98 additions~\citep{Laderman1976ANA}. Recent advances have dramatically lowered this count: Schwartz and Vaknin achieved 61 additions using a change of basis~\citep{schwartz2023pebbling}, Mårtensson and Wagner achieved 62~\citep{maartensson2025number}, and Stapleton recently set a new record of 60 additions without a change of basis \citep{stapleton202560}.

This work reports a further reduction to 58 additions, achieved without a change of basis. The result is obtained through a parallel heuristic search algorithm based on an extension of the flip-graph approach with ternary restrictions~\citep{perminov2025fast} and a greedy intersection reduction strategy~\citep{perminov2025parallel}. The main result is a new rank-23 scheme using only 58 additions and employing exclusively ternary coefficients $\{-1, 0, 1\}$. This ensures the scheme's portability and efficiency across arbitrary fields.

\section{Methodology}

\subsection{Matrix Multiplication Schemes}
\label{subsec:schemes}

A matrix multiplication scheme for multiplying matrices $A \in \mathbb{F}^{n \times n}$ and $B \in \mathbb{F}^{n \times n}$ over a field $\mathbb{F}$ with rank $r$ is defined by three coefficient tensors $U \in \mathbb{F}^{r \times n \times n}$, $V \in \mathbb{F}^{r \times n \times n}$ and $W \in \mathbb{F}^{r \times n \times n}$. It computes $r$ intermediate scalar products:

\begin{align*}
m_1 = (u^{(1)}_{11} a_{11} + \dots + u^{(1)}_{nn} a_{nn})\;\cdot\; & (v^{(1)}_{11} b_{11} + \dots + v^{(1)}_{nn} b_{nn})\\
\vdots\;  \\
m_r = (u^{(r)}_{11} a_{11} + \dots + u^{(r)}_{nn} a_{nn})\;\cdot\; & (v^{(r)}_{11} b_{11} + \dots + v^{(r)}_{nn} b_{nn}),
\end{align*}

and reconstructs the result matrix $C$, where $C = AB$, as:

\begin{align*}
c_{ij} = w^{(1)}_{ij}m_1 + \dots + w^{(r)}_{ij}m_r.
\end{align*}

The tensors $U, V, W$ must satisfy the Brent equations~\citep{brent1970algorithms}. This work focuses on the $n=3$, $r=23$ case where all coefficients in $U, V, W$ are restricted to the set $\{-1, 0, 1\}$.

\subsection{Ternary Flip-Graph Search}
\label{subsec:flip_graph}
The search operates on the space of valid 3×3 multiplication schemes of varying rank. This space is navigated using an extension of the flip-graph approach~\citep{kauers2023flip}, constrained by ternary restrictions~\citep{perminov2025fast}. A flip is a local transformation applied to the tensors $U, V, W$. It modifies the coefficients while preserving the correctness of the scheme (i.e., satisfaction of the Brent equations) and maintaining its current rank. Flips enable traversal between different schemes with equal rank.

The ternary restriction enforces that all coefficients in $U, V, W$ remain within the set $\{-1, 0, 1\}$ throughout the search. This constraint focuses the exploration on schemes that are directly implementable over any field. Many existing flip-graph methods~\citep{kauers2023flip, moosbauer2025flip} first search for schemes over the binary field $\mathbb{Z}_2$, which is computationally efficient, and later attempt to "lift" the found coefficients to integer or rational values to ensure validity over rings such as $\mathbb{Z}$ or $\mathbb{Q}$. The ternary approach avoids this lifting step entirely by working directly with the target coefficient set.

The core discovery process is a random walk on this ternary flip graph. To escape irreducible states where no applicable flip can improve the scheme, a plus operator is used. This operator, defined in~\citep{arai2024adaptive}, increases the rank of a scheme by one while preserving correctness and the ternary coefficient property. After its application, the flip-graph walk continues, often enabling further optimization that can later reduce the rank back to the target value.

\subsection{Greedy Intersection Reduction}
\label{subsec:reduction}

After obtaining a candidate rank-23 scheme, its additive complexity is minimized using a common subexpression elimination (CSE) algorithm \cite{perminov2025parallel}. During the main search loop, the CSE step uses a single heuristic randomly chosen from a collection of strategies, including the Greedy-Intersections method.

The algorithm identifies all candidate subexpressions of the form $x_i \pm x_j$ within the linear forms defined by the $U$, $V$ and $W$ tensors. A heuristic scoring function evaluates every identified candidate. The subexpression with the highest score is selected, replaced by a fresh variable, and eliminated throughout the entire scheme.

After each substitution, the algorithm identifies all new candidate subexpressions in the updated scheme and selects the next best one. The process continues until no remaining candidate subexpression would reduce the total addition count.

\section{Discovery Algorithm}
\label{subsec:algorithm}

The discovery algorithm integrates the ternary flip-graph search and greedy intersection reduction into a three-phase iterative process. The search begins by initializing a naive scheme for 3×3 matrix multiplication. This naive scheme has a rank of 27 and uses only coefficients from $\{0, 1\}$ (the standard matrix multiplication algorithm). The core procedure then executes the continuous loop described in Algorithm~\ref{alg:search}.

\begin{algorithm}[H]
\caption{Search loop for low-addition schemes}
\label{alg:search}
\KwIn{a naive 3×3 multiplication scheme $S$ with coefficients $\{0,1\}$ and rank $27$.}
\KwOut{a scheme $S$ with minimal addition count discovered during search.}
\BlankLine
\While{search not manually stopped}{
    \tcp{flip to target rank}
    \While{$rank(S) \neq 23$}{
        $flip(S)$\;
        \BlankLine
        \If{no flip is possible \Or $random() < 0.05$}{
            $plus(S)$\;
        }
    }
    \BlankLine
    \tcp{reduce additions with greedy intersection}
    Execute greedy intersection reduction algorithm on $S$\;
    Check for improvements in addition count\;
    \BlankLine
    \tcp{random change}
    $flip(S)$\;
    \BlankLine
    \While{$rank(S) < 25$ \And $random() < 0.05$}{
        $plus(S)$\;
    }
}
\end{algorithm}

\textbf{Phase 1 (flip to target rank):} the algorithm uses random flips to change the scheme until it reaches rank 23. Each flip converts the scheme to another scheme. If a flip creates a tensor component with all zero coefficients, the rank of the scheme is reduced by one. If no useful flip is possible, or with a 5\% chance even if a flip is possible, the plus operator is used. This operator increases the rank by one, which helps to continue the search when stuck.

\textbf{Phase 2 (reduce additions with greedy intersection):} after reaching rank 23, the greedy intersection reduction algorithm runs. It finds common subexpressions in the formulas from the $U$, $V$ and $W$ tensors. These common parts are replaced with new variables. This step repeats, choosing the best replacement each time, to lower the total number of additions.

\textbf{Phase 3 (random change):} to keep searching new options and avoid getting stuck, a random flip is applied to change the scheme a bit. Also, with a 5\% chance, the plus operator is used if the current rank is less than 25. This creates small changes that can lead to better schemes later.

The three phases execute in a continuous loop. The search runs indefinitely until manually stopped, allowing for exhaustive exploration of the solution space. This design balances targeted optimization (Phase 2) with broader exploration (Phases 1 and 3) to systematically navigate toward schemes with minimal addition counts.
\clearpage

\section{Discovered 58-Addition Scheme}

The discovered algorithm for exact 3×3 matrix multiplication is presented below. It requires 23 multiplications and achieves an additive complexity of 58. The scheme was derived from an initial, unreduced algorithm requiring 120 naive additions. Through common subexpression elimination, 20 fresh intermediate variables were introduced: 4 for linear combinations of matrix $A$ ($u_1$–$u_4$), 8 for matrix $B$ ($v_1$–$v_8$), and 8 for combining the products ($w_1$–$w_8$). The complete algorithm is constructed from these intermediates, which are computed once and reused. A notable feature of the scheme is that all expressions are defined without variable inversions (the first non-zero coefficient in each defining equation is $+1$). The 58 additive operations consist of 34 additions and 24 subtractions.

\begin{align*}
    & A = \begin{bmatrix}
        a_{11} & a_{12} & a_{13} \\
        a_{21} & a_{22} & a_{23} \\
        a_{31} & a_{32} & a_{33} \\
    \end{bmatrix}
    &&
    B = \begin{bmatrix}
        b_{11} & b_{12} & b_{13} \\
        b_{21} & b_{22} & b_{23} \\
        b_{31} & b_{32} & b_{33} \\
    \end{bmatrix}
\\
\\
& u_{1} = a_{31} + a_{33} && m_{1} = u_{1} \times v_{5} \\
& u_{2} = a_{21} + a_{22} && m_{2} = u_{2} \times (v_{2} + v_{6}) \\
& u_{3} = a_{13} + u_{1} && m_{3} = a_{32} \times b_{23} \\
& u_{4} = a_{32} - u_{2} && m_{4} = a_{31} \times (b_{13} + v_{7}) \\
&  && m_{5} = u_{3} \times v_{7} \\
& v_{1} = b_{22} + b_{32} && m_{6} = (a_{32} - a_{33}) \times b_{22} \\
& v_{2} = b_{31} - v_{1} && m_{7} = a_{23} \times b_{33} \\
& v_{3} = b_{12} + v_{2} && m_{8} = (u_{1} - u_{2}) \times v_{2} \\
& v_{4} = b_{11} - v_{3} && m_{9} = (a_{12} - a_{13}) \times b_{22} \\
& v_{5} = b_{33} + v_{1} && m_{10} = (u_{3} - a_{21}) \times v_{3} \\
& v_{6} = b_{21} - b_{23} && m_{11} = (a_{13} + a_{33}) \times (v_{1} - b_{12}) \\
& v_{7} = b_{12} - v_{5} && m_{12} = a_{13} \times b_{33} \\
& v_{8} = v_{4} - v_{6} && m_{13} = (a_{31} + u_{4}) \times v_{4} \\
&  && m_{14} = a_{11} \times b_{13} \\
&  && m_{15} = (a_{11} + u_{3}) \times b_{12} \\
& w_{1} = m_{5} + m_{12} && m_{16} = (a_{13} + a_{23}) \times b_{31} \\
& w_{2} = m_{1} + w_{1} && m_{17} = (a_{22} - a_{32}) \times (v_{4} - b_{21}) \\
& w_{3} = m_{8} + w_{2} && m_{18} = (a_{11} + a_{21}) \times b_{11} \\
& w_{4} = m_{3} - m_{23} && m_{19} = a_{12} \times b_{23} \\
& w_{5} = m_{2} + w_{4} && m_{20} = (a_{12} + a_{22}) \times b_{21} \\
& w_{6} = w_{3} + w_{5} && m_{21} = a_{21} \times (b_{13} - v_{8}) \\
& w_{7} = m_{10} - w_{6} && m_{22} = (a_{22} - a_{23}) \times (b_{31} - b_{32}) \\
& w_{8} = m_{17} + w_{7} && m_{23} = u_{4} \times v_{8} \\
\end{align*}

\[
C = 
    \begin{bmatrix}
m_{18} + m_{20} + w_{8} && m_{9} + m_{15} - w_{2} && m_{12} + m_{14} + m_{19} \\
m_{16} - w_{8} && m_{16} + m_{22} + w_{3} - m_{10} && m_{7} + m_{21} + w_{4} - m_{17} \\
m_{11} + m_{13} + w_{6} && m_{6} + m_{11} + w_{2} && m_{3} + m_{4} - m_{11} - w_{1} \\
    \end{bmatrix}
\]

\section{Discussion}
\label{sec:discussion}

The discovery of a 58-addition scheme advances the state of the art for practical 3×3 matrix multiplication. The search that found this scheme required approximately 30 minutes of computation on a standard laptop CPU (Intel Core i7-9750H), demonstrating that state-of-the-art results can be achieved without specialized high-performance computing resources. Reducing the addition count from 60 to 58 directly lowers the operation cost in performance-critical applications like computer graphics and scientific computing, where such multiplications are performed repeatedly. The scheme's use of only ternary coefficients $\{-1, 0, 1\}$ is a significant practical strength, ensuring it consists solely of cheap additions and subtractions, making it portable and efficient across different hardware and numerical fields.

This result validates the effectiveness of the combined flip-graph and heuristic search methodology. It demonstrates that focused, resource-efficient strategies can achieve top results in this domain. The success stems from two key choices: restricting the search to ternary coefficients to maintain practicality, and applying a greedy heuristic capable of finding deep optimizations in additive complexity. The progression from Laderman's 98 additions to 58 shows that substantial optimization remains possible even for long-studied algorithms.

Looking forward, it remains an open question whether 58 additions is the minimum possible for a rank-23 scheme. The presented methodology can be applied to search for optimizations in other small matrix formats. A valuable next step would be to implement this scheme in a production BLAS library to measure its real-world performance improvement.

\section{Conclusion}
\label{sec:conclusion}

This work presented a new algorithm for 3×3 matrix multiplication that achieves a rank of 23 with only 58 scalar additions. This result improves the previous state-of-the-art additive complexity of 60 additions, established without a change of basis. Combined with the 23 essential multiplications, the total number of scalar operations is reduced from 83 to 81. The scheme uses coefficients exclusively from the set $\{-1, 0, 1\}$, ensuring efficiency and portability across hardware architectures and numerical fields.

The scheme was discovered through an automated search methodology that combines a ternary-restricted flip-graph walk with a greedy intersection reduction heuristic. This approach demonstrates that effective exploration of the algorithm space can yield significant improvements through structured combinatorial search.

The new scheme offers a direct path to performance gains in low-level numerical libraries. Future work may focus on establishing a lower bound for the additive complexity of rank-23 schemes, applying the search methodology to other matrix formats, and implementing this algorithm in production BLAS software for practical benchmarking.

\appendix
\section{Python verification Script}
\label{app:verification}

The scheme was verified through symbolic and numerical methods. Symbolically, all intermediate variables were substituted to obtain the original linear forms. This process reconstructed the full coefficient tensors $U$, $V$ and $W$, which were confirmed to satisfy the Brent equations. This proves the algorithm's correctness for all $3 \times 3$ matrices over any field.

Numerically, a Python script executed the algorithm on 10,000 random matrix pairs. All results matched standard multiplication in each test.

\begin{lstlisting}[style=py,caption={Python script for numerical validation of the 58-addition scheme}]
import numpy as np

def multiply_3x3_rank23_58additions(a: np.ndarray, b: np.ndarray) -> np.ndarray:
    a11, a12, a13, a21, a22, a23, a31, a32, a33 = a.ravel()
    b11, b12, b13, b21, b22, b23, b31, b32, b33 = b.ravel()

    u1 = a31 + a33
    u2 = a21 + a22
    u3 = a13 + u1
    u4 = a32 - u2

    v1 = b22 + b32
    v2 = b31 - v1
    v3 = b12 + v2
    v4 = b11 - v3
    v5 = b33 + v1
    v6 = b21 - b23
    v7 = b12 - v5
    v8 = v4 - v6

    m1 = u1 * v5
    m2 = u2 * (v2 + v6)
    m3 = a32 * b23
    m4 = a31 * (b13 + v7)
    m5 = u3 * v7
    m6 = (a32 - a33) * b22
    m7 = a23 * b33
    m8 = (u1 - u2) * v2
    m9 = (a12 - a13) * b22
    m10 = (u3 - a21) * v3
    m11 = (a13 + a33) * (v1 - b12)
    m12 = a13 * b33
    m13 = (a31 + u4) * v4
    m14 = a11 * b13
    m15 = (a11 + u3) * b12
    m16 = (a13 + a23) * b31
    m17 = (a22 - a32) * (v4 - b21)
    m18 = (a11 + a21) * b11
    m19 = a12 * b23
    m20 = (a12 + a22) * b21
    m21 = a21 * (b13 - v8)
    m22 = (a22 - a23) * (b31 - b32)
    m23 = u4 * v8

    w1 = m5 + m12
    w2 = m1 + w1
    w3 = m8 + w2
    w4 = m3 - m23
    w5 = m2 + w4
    w6 = w3 + w5
    w7 = m10 - w6
    w8 = m17 + w7

    c11 = m18 + m20 + w8
    c12 = m9 + m15 - w2
    c13 = m12 + m14 + m19
    c21 = m16 - w8
    c22 = m16 + m22 + w3 - m10
    c23 = m7 + m21 + w4 - m17
    c31 = m11 + m13 + w6
    c32 = m6 + m11 + w2
    c33 = m3 + m4 - m11 - w1

    return np.array([[c11, c12, c13], [c21, c22, c23], [c31, c32, c33]])


def validate(n: int):
    for _ in range(n):
        a = np.random.randint(-100, 101, size=(3, 3))
        b = np.random.randint(-100, 101, size=(3, 3))
        assert np.array_equal(multiply_3x3_rank23_58additions(a, b), a @ b)

    print("Tests passed")


if __name__ == "__main__":
    validate(10000)
\end{lstlisting}

\bibliographystyle{unsrtnat}
\bibliography{references}  %%% Uncomment this line and comment out the ``thebibliography'' section below to use the external .bib file (using bibtex) .





% \end{thebibliography}

\end{document}